\documentclass[journal]{IEEEtran}
\usepackage{amsmath}
\usepackage{amsfonts}
\usepackage{epsfig}
\usepackage{amssymb}
\usepackage{graphics}
\usepackage{}
\usepackage{float}
\usepackage{amsmath}
\usepackage{slashbox}
\usepackage[T1]{fontenc}
\usepackage{cite}
\usepackage{cleveref}
\usepackage{array}
\usepackage{caption}
\usepackage{subcaption}
\usepackage[usenames, dvipsnames]{color}

\makeatletter
\newcommand{\thickhline}{%
    \noalign {\ifnum 0=`}\fi \hrule height 1.2pt
    \futurelet \reserved@a \@xhline
}
\newcolumntype{"}{@{\hskip\tabcolsep\vrule width 1pt\hskip\tabcolsep}}
\makeatother

\begin{document}

\title{Modeling and Mitigating Errors in Belief Propagation for Distributed Detection}
%\title{Robust Belief Propagation via Linear Data Fusion: A Distributed Detection Scenario}

\author{\authorblockN{Younes~Abdi,~\IEEEmembership{Member,~IEEE,} and Tapani~Ristaniemi,~\IEEEmembership{Senior Member,~IEEE}}
\thanks{Y.~Abdi and T.~Ristaniemi are with the Faculty of Information Technology, University of Jyv\"askyl\"a, P.~O.~Box 35, FIN-40014, Jyv\"askyl\"a, Finland, Tel. +358 40 7214 218 \mbox{(e-mail:younes.abdi@jyu.fi, tapani.ristaniemi@jyu.fi)}. 
}\vspace{-0.25 in}
}

\maketitle

\begin{abstract}
We study the behavior of the belief-propagation (BP) algorithm affected by erroneous data exchange in a wireless sensor network (WSN). The WSN conducts a distributed binary hypothesis test where the joint statistical behavior of the sensor observations is modeled by a Markov random field whose parameters are used to build the BP messages exchanged between the sensing nodes. Through linearization of the BP message-update rule, we analyze the behavior of the resulting erroneous decision variables and derive closed-form relationships that describe the impact of stochastic errors on the performance of the BP algorithm. We then develop a decentralized distributed optimization framework to enhance the system performance by mitigating the impact of errors via a distributed linear data-fusion scheme. Finally, we compare the results of the proposed analysis with the existing works and visualize, via computer simulations, the performance gain obtained by the proposed optimization. 
\end{abstract}


\section{Introduction}\label{sec:Intro}
\PARstart{D}{ealing} with a large collection of random variables and their interactions is a common practice when designing statistical inference systems. Graphical models, a.k.a., factor graphs, which are commonly used to capture the interdependencies between correlated random variables, are known to provide a powerful framework for developing effective low-complexity inference algorithms in various fields such as wireless communications, image processing, combinatorial optimization, and machine learning, see e.g.,  \cite{Kschischang01, Noorshams13, Loeliger04}. Belief propagation (BP) \cite{Wainwright08} is a well-known statistical inference algorithm that works based on parallel message-passing between the nodes in a factor graph. BP is sometimes referred to as the sum-product algorithm.  

When working with the BP algorithm, we should bear in mind that digital computation and digital communication are both error-prone processes in general. The messages exchanged between the nodes in a wireless network can always be adversely affected by errors caused by unreliable hardware components, quantization processes, approximate representations, wireless channel impairments, etc. Even though the BP algorithm has been extensively studied in the literature, we have rather limited knowledge about how stochastic errors in messages affect the beliefs obtained and how these erroneous beliefs influence the result of statistical inference schemes implemented by the BP algorithm. This territory is difficult to explore mainly due to the nonlinearities in the BP message-passing iteration. 

In \cite{Abdi20}, we have developed a systematic framework for analyzing the behavior of BP and optimizing its performance in a distributed detection scenario. In particular, we have shown that the decision variables built by the BP algorithm are, approximately, \emph{linear} combinations of the local likelihoods in the network. Consequently, we have derived in \cite{Abdi20} closed-form relationships for the system performance metrics and formulated a distributed optimization scheme to achieve a near-optimal detection  performance. Moreover, we have discussed the relationship between the BP and the max-product algorithms in \cite{Abdi2019arxiv} where we extend the proposed framework in \cite{Abdi20} to optimize the performance of the max-product algorithm in a distributed detection scenario. In this paper, we further extend that framework to gain insight into the impact of computation and communication errors, in a BP iteration, on the resulting decision variables and to effectively mitigate that impact. Examples of BP being used in distributed detection can be found in \cite{Penna12, Wymeersch12, Li10, Zhang11}.

Accumulation of message errors and their adverse effect on the performance of BP is analyzed in \cite{Ihler05} where the message errors are modeled as uncorrelated random variables to find probabilistic guarantees on the magnitude of errors affecting the beliefs. Assuming uncorrelated behavior for errors is inspired in \cite{Ihler05} by observing the behavior and stability of digital filters, in the presence of quantization effects, which can be analyzed reliably by assuming uncorrelated behavior in the corresponding random errors \cite{Willsky78}. Such a modeling approach is in line with the von Neumann model of noisy circuits \cite{von56}, which considers transient faults in logic gates and wires as message and node computation noise that is both spatially and temporally independent \cite{Varshney11}. 

The behavior of BP implemented on noisy hardware is investigated in \cite{Huang15} where it is observed that under the so-called \emph{contracting mapping condition} \cite{Mooij07}, the distance between successive messages in a noise-free BP decreases by the number of iterations. Consequently, in the presence of hardware (or computation) noise, the faulty messages which violate this trend can be detected and discarded (censored) from the BP iterations. Such an approach is termed \emph{censoring BP} in \cite{Huang15} and is shown to performs well when the hardware noise distribution has a large mass at zero and has non-negligible masses at some points sufficiently away from zero. As an alternative approach, the so-called \emph{averaging BP} (ABP) is also proposed in \cite{Huang15}. In this method, as the name implies, an average of the messages up to the last iteration is saved and then used, instead of the actual messages, to build the beliefs. This method is proposed and its convergence is established for general zero-mean computation noise distributions. Again, the von Neumann model is used in \cite{Huang15} to analyze the behavior of message errors. 

  In this paper, we use the fact that the BP algorithm and the linear data-fusion scheme are elegantly related to each other in the context of distributed detection. Fortunately, there already exists a rich collection of scientific works in the literature that investigate low-complexity detector structures based on linear fusion in various design scenarios \cite{Quan08, Quan09, Quan10, Taricco11, Abdi14}. In many of these works, the data-exchange process within the sensor network is assumed adversely affected by non-idealities in the underlying communication links. Hence, dealing with erroneous data is a familiar challenge when designing wireless sensor networks (WSN). We use this knowledge to cope with the impact of message errors on distributed detection systems realized by BP. 
  
In particular, we approximate the message structure in BP by a linear expression to study the impact of erroneous data exchange on the BP algorithm and to clarify how it affects the performance of the distributed detection concerned. We derive approximate expressions to measure the strength of the cumulative errors that affect the BP-based decision variables. These expressions are in the form of mean-squared error (MSE) levels. We compare the MSE levels obtained with the one in \cite{Ihler05} to gain insight into the behavior of BP and to see how computation and communication errors propagate throughout the underlying factor graph. Moreover, based on the proposed linear approximation, we show that ABP is effective in alleviating message errors and falls short of mitigating the impact of erroneous local likelihood ratios (LLRs) on the resulting decision variables. 

We also show, under practical assumptions, that the decision variables built by an erroneous BP are disturbed by a sum of independent error components whose collective impact can be modeled, approximately, by Gaussian random variables. Consequently, we establish the probability distribution of the resulting erroneous decision variables, derive the performance metrics of the BP-based distributed detection in closed form, and propose a two-stage optimal linear fusion scheme to cope with the impact of errors on the system performance. We then develop a blind adaptation algorithm to realize the proposed two-stage optimization when the statistics describing the radio environment are not available \emph{a priori}.  

Here is an overview of the paper organization: In Sec. \ref{sec:linFusBP}, we briefly explain the use of linear fusion and BP in distributed detection and provide the related formulations. In Sec. \ref{sec:errorsInBP}, we discuss errors in BP and model their impact on the decision variables obtained. In Sec. \ref{sec:MitigLinFus}, we view BP as a distributed linear fusion and formulate the proposed optimization framework. In Sec. \ref{sec:simulations}, we conduct computer simulations to verify our analysis and to illustrate how effectively the proposed method mitigates the impact of errors in a WSN with faulty devices. Finally, we provide our concluding remarks in Sec. \ref{sec:conclusions}. 

\section{Linear Fusion and Belief Propagation for Distributed Detection} \label{sec:linFusBP}
We consider $N$ binary random variables, represented by $\boldsymbol{x} = [x_1, ..., x_N]^T$, whose status are estimated based on $N$ observations denoted $ \boldsymbol{Y} = [\boldsymbol{y}_1, ..., \boldsymbol{y}_N]$ made by a network of $N$ sensing nodes. Each node, say node $i$, which intends to estimate the status of $x_i$, collects $K$ observation samples, denoted by $\boldsymbol{y}_i = [y_i(1), ..., y_i(K)]^T$, and exchanges information with other nodes in the network to realize together a binary hypothesis test as  $\hat{\boldsymbol{x}} = \operatorname*{max}_{\boldsymbol{x}} p(\boldsymbol{x}\vert \boldsymbol{Y}) = \operatorname*{max}_{\boldsymbol{x}} p(\boldsymbol{Y}\vert \boldsymbol{x})p(\boldsymbol{x})$. 
This test can be conducted with low implementation complexity in two alternative ways that are explained in the following. 

\subsection{Linear Data-Fusion} \label{sec:linearFusion}
Linear fusion has been extensively used in the context of spectrum sensing where the aim is to detect the presence or absence of a target signal by evaluating noisy observations made throughout a wireless sensor network (WSN). For brevity, we explain the uni-variate case here. In this detection scenario, each node, say node $i$, collects the signal samples $\boldsymbol{y}_i = x\boldsymbol{s}_i + \boldsymbol{n}_i$, where the random variable $x \in \{0,1\}$ determines the presence or absence of the target signal $\boldsymbol{s}_i$ in the radio environment. In this model, $\boldsymbol{n}_i $ denotes a vector of zero-mean Gaussian white noise samples while $\boldsymbol{s}_i$ denotes an unknown deterministic vector of target signal samples, which, in general, represent a superposition of multiple signals received at node $i$ from different transmitters. %The sensing nodes \emph{report} the data they collect, i.e., $\boldsymbol{y}_i$'s, to a so-called \emph{fusion center} where the detection is performed. 

The optimal approach to such a detection is known to be the so-called \emph{likelihood-ratio test} (LRT) \cite{Kay93}, which is conducted by evaluating the LLR, i.e., by $\hat{x} = \mathbf{1}\{\lambda_{\textup{LRT}}\}$ where 
\begin{equation} \label{eq:lambda_LRT}
\lambda_{\textup{LRT}} \triangleq \ln \frac{p(\boldsymbol{Y}|x = 1)}{p(\boldsymbol{Y}|x = 0)} = \sum_{i=1}^{N} \gamma_i
\end{equation}
where 
\begin{equation} \label{eq:matchFilt}
\gamma_i \triangleq \ln \frac{p(\boldsymbol{y}_i|x = 1)}{p(\boldsymbol{y}_i|x = 0)}  = \boldsymbol{s}_{i}^T \boldsymbol{y}_{i} - \frac{1}{2} \|\boldsymbol{s}_{i}\|^2
\end{equation}
where $\gamma_i$ is referred to as the \emph{local} LLR at node $i$. 
By $\mathbf{1}\{\cdot\}$ we represent the indicator function that returns one if its argument is positive and returns zero otherwise.  It is clear that the LRT is, in fact, a \emph{matched-filtering} process, which requires the target signal to be known \emph{a priori} at the sensing nodes. 
In practice, the local sensing process is realized by \emph{energy detection} due to its ease of implementation and due to the fact that its structure does not depend on the behavior of the target signal. Energy detection is realized by $\gamma_i \triangleq  \frac{1}{K} \|\boldsymbol{y}_{i}\|^2$ and the sensor outcomes are combined linearly to build a \emph{global} test statistic \cite{Quan08, Quan09, Quan10, Taricco11, Abdi14}, i.e.,  
\begin{equation} \label{eq:lambda_LF}
\lambda_{\textup{LF}} \triangleq \sum_{i=1}^{N} w_i\gamma_i = \boldsymbol{w}^T \boldsymbol{\gamma}
\end{equation}
where $\boldsymbol{w} \triangleq [w_1, ..., w_N]^T$ and $\boldsymbol{\gamma} \triangleq [\gamma_1, ..., \gamma_N]^T$. Then, $\lambda_{\textup{LF}}$ is compared against a predefined threshold $\tau$ to conduct the hypothesis test, i.e., $\hat{x} = \mathbf{1}\{\lambda_{\textup{LF}} > \tau\}$. Assuming that the number of signal samples $K$ is large enough \cite{Quan08, Quan09, Quan10, Taricco11, Abdi14} such that $\gamma_i$'s in \eqref{eq:lambda_LF} behave as Gaussian random variables, we can model the test summary $\lambda_{\textup{LF}}$, given the status of $x$, as a Gaussian random variable as well. Specifically, for $x = b$ we have $\lambda_{\textup{LF}} \sim \mathcal{N} \left (\boldsymbol{w}^T\boldsymbol{\mu}_b, \boldsymbol{w}^T\boldsymbol{\Sigma}_b \boldsymbol{w} \right) $ where for $b \in \{0,1\}$, $\boldsymbol{\mu}_b \triangleq \textup{E}[\boldsymbol{\gamma} | x = b]$ and $\boldsymbol{\Sigma}_b \triangleq \textup{cov}(\boldsymbol{\gamma} | x = b)$. Consequently, the false-alarm and detection probabilities, denoted $P_\textup{f}$ and $P_\textup{d}$ respectively, of this detector are derived in closed form by
\begin{equation} \label{eq:g}
g_b(\tau, \boldsymbol{w}) \triangleq \textup{Pr}\{\lambda_{\textup{LF}} > \tau | x = b\} = Q\left ( \frac{\tau - \boldsymbol{w}^T\boldsymbol{\mu}_b}{\sqrt{\boldsymbol{w}^T\boldsymbol{\Sigma}_b \boldsymbol{w}}}  \right )
\end{equation}
where $Q(\cdot)$ denotes the $Q$-function. Note that $P_\textup{f} = g_0(\tau, \boldsymbol{w})$ while $P_\textup{d} = g_1(\tau, \boldsymbol{w})$. By setting $\tau = Q^{-1}(\alpha)\boldsymbol{w}^T\boldsymbol{\Sigma}_0 \boldsymbol{w} + \boldsymbol{w}^T\boldsymbol{\mu}_0$, we have $P_\textup{f} = \alpha$ and then the detection performance can be optimized by maximizing the resulting $P_\textup{d}$ over $\boldsymbol{w}$. This is the well-known \emph{Neyman-Pearson} approach \cite{Kay93}. Through some algebraic manipulations, this optimization is formally stated as  
\begin{equation} \label{eq:w_opt}
\boldsymbol{w}^* = \arg\min_{\boldsymbol{w}}  \frac{Q^{-1}(\alpha)\sqrt{\boldsymbol{w}^T\boldsymbol{\Sigma}_0 \boldsymbol{w}} - \boldsymbol{w}^T\boldsymbol{\delta}}{\sqrt{\boldsymbol{w}^T\boldsymbol{\Sigma}_1 \boldsymbol{w}}}  
\end{equation}
where $\boldsymbol{\delta} \triangleq \boldsymbol{\mu}_1 - \boldsymbol{\mu}_0$. This problem is solved by quadratic programming in \cite{Quan08}, by semidefinite programming in \cite{Quan10}, and by invoking the Karush-Kuhn-Tucker conditions in \cite{Taricco11}. From these works, we know that the performance of linear fusion is close to the LRT performance. Alternatively, we can maximize the so-called \emph{deflection coefficient} of the detector. This approach, which has a low computational complexity and leads to a good performance level, is realized by  
\begin{equation} \label{eq:w_DC}
\boldsymbol{w}^* =\arg\max_{\boldsymbol{w}} \begin{matrix}
  \Delta^2(\boldsymbol{w}), & \textup{s.t.,}~  \left \| \boldsymbol{w} \right \| = 1
\end{matrix} 
\end{equation}
 where  
\begin{equation} \label{eq:DC}
\Delta^2(\boldsymbol{w}) \triangleq \frac{\left (\textup{E}[\lambda_{\textup{LF}}|x = 1]-\textup{E}[\lambda_{\textup{LF}} |x = 0]  \right )^2 }{\textup{Var}[\lambda_{\textup{LF}}|x = 0]} = \frac{\left (\boldsymbol{w}^T\boldsymbol{\delta}  \right )^2}{\boldsymbol{w}^T\boldsymbol{\Sigma}_{0}\boldsymbol{w}}
\end{equation}
Consequently, by using the Rayleigh-Ritz inequality \cite{Quan08}, $\boldsymbol{w}^*$ is obtained in closed form as $ \boldsymbol{w}^* = \boldsymbol{\Sigma}_{0}^{-1}\boldsymbol{\delta}/\left \| \boldsymbol{\Sigma}_{0}^{-1}\boldsymbol{\delta} \right \|$. When $\boldsymbol{\Sigma}_{1}$ is used in \eqref{eq:DC}, the objective function is referred to as \emph{modified deflection coefficient}. 

Note that both optimizations in \eqref{eq:w_opt} and \eqref{eq:w_DC} can be realized while taking into account the impact of erroneous $\gamma_i$'s. The so-called \emph{reporting errors} in \cite{Quan08, Quan09, Quan10, Taricco11, Abdi14} model the impact of erroneous communication links through which the sensing nodes share their observations. The optimal fusion weights obtained by \eqref{eq:w_opt} and \eqref{eq:w_DC} emphasize the impact of local sensing outcomes generated in high SNR conditions while suppressing the impact of errors caused by the data communication process between the sensing nodes. 

Extension of the linear detection structure in \eqref{eq:lambda_LF} to $N$ variables is discussed in \cite{Quan09}, in the context of multiband spectrum sensing, where the detection performance is optimized by the so-called \emph{sequential optimization} method that is based on maximizing the deflection coefficient of the system. In the following, we discuss the BP algorithm and show that it can be interpreted as a multivariate linear data-fusion. 

\subsection{Belief Propagation}
We model the sensor network structure concerned by a Markov random field (MRF) defined on an undirected graph $G = (\mathcal{V,E})$. In this model, the set of vertices $\mathcal{V}$ corresponds to the set of network nodes while each edge $(i,j) \in \mathcal{E}$  represents a possible connection between nodes $i$ and $j$. Each node, say node $i$, is associated with a random variable $x_i$ and the edge $(i,j)$ models a possible correlation between $x_i$ and $x_j$. This model fits well into the commonly-used ad-hoc network configurations in which major network functionalities are conducted through pairwise i.e., one-hop, links between the nodes located close to each other. This design method is based on the common assumption that nodes located close enough to each other for one-hop communication, experience some levels of correlation between their sensor outcomes.

By using the MRF, we write $p({\boldsymbol x}\vert {\boldsymbol Y}) $ as a product of univariate and bivariate functions, i.e.,    
\begin{equation}\label{eq:MRF} 
p({\boldsymbol x}\vert {\boldsymbol Y}) \propto  \prod_{n=1}^{K} \phi_{n}(x_{n}) \prod_{(i,j) \in {\cal E}} \psi_{ij}(x_i, x_j) 
\end{equation}
Note that $\propto$ in \eqref{eq:MRF} refers to a normalization that ensures $\sum_{\boldsymbol x}p({\boldsymbol x}\vert {\boldsymbol Y}) = 1 $. When including the bivariate terms in the product, each edge in the factor graph is included in the product only once. This is realized by doing the multiplication on $i < j$ while $i \in \mathcal{N}_j$. We use $\mathcal{N}_j$ to denote the set of neighbors of node $j$ in the graph, i.e., $\mathcal{N}_j \triangleq \{k : (k,j) \in \mathcal{E}\}$. By using \eqref{eq:MRF}, we formulate the message received at node $j$ from node $k$ as 
\begin{equation}\label{mu_kj}
\mu_{k \rightarrow j}^{(l)}(x_j) \propto \sum_{x_k} \phi_k(x_k)\psi_{kj}(x_k,x_j) \prod_{n \in \mathcal{N}_k^j}\mu_{n \rightarrow k}^{(l-1)}(x_k)
\end{equation}
 where by $\mathcal{N}_k^j \triangleq \mathcal{N}_k \backslash \{j\}$ we  denote all nodes connected to node $k$ except for node $j$. We denote by $b_j^{(l)}(x_j)$ the belief, about the status of $x_j$, formed at node $j$, which is obtained via multiplying the potential at node $j$ by the messages received from all its neighbors, i.e.,  
\begin{equation}\label{b_j} 
b_j^{(l)}(x_j) \propto \phi_{j}(x_j) \prod_{k \in \mathcal{N}_j}\mu_{k \rightarrow j}^{(l)} (x_j)
\end{equation}
The beliefs are used as estimates of the desired marginal distributions, i.e., $ b_j^{(l)}(x_j) \approx p(x_j \vert \boldsymbol{Y})$. 
By adopting the commonly-used exponential model \cite{Wainwright08} to represent the \emph{a priori} probability measure defined on $\boldsymbol{x}$, we have
 \begin{equation}\label{eq:p(x)} 
p({\boldsymbol x}) \propto  \textup{exp}\left ( \sum_{n=1}^{K} \theta_{n}x_{n} + \sum_{(i,j) \in {\cal E}} J_{ij}x_i x_j \right )
\end{equation}
For a given ${\boldsymbol x}$, we assume the local observations to be mutually independent. Consequently, we have \cite{Abdi20} 
\begin{equation}\label{eq:p(x|y)_2} 
p({\boldsymbol x}\vert {\boldsymbol Y}) \propto \prod_{k=1}^{N} p(\boldsymbol{y}_k \vert x_k) e^{\theta_k x_k} \prod_{(i,j) \in \mathcal{E} } e^{J_{ij}x_ix_j}
\end{equation}
Hence, by using \eqref{eq:p(x|y)_2}, the BP messages are built as 
\begin{equation}\label{mu_kjExp} 
\mu_{k \rightarrow j}^{(l)} (x_j) \propto \sum_{x_k} p(\boldsymbol{y}_k \vert x_k) e^{\theta_kx_k} e^{J_{kj}x_kx_j} \prod_{n \in \mathcal{N}_k^j}\mu_{n \rightarrow k}^{(l-1)} (x_k)
\end{equation}
and the beliefs at iteration $l$ are expressed as 
\begin{equation}\label{b_jExp} 
b_j^{(l)}(x_j) \propto p(\boldsymbol{y}_j \vert x_j) e^{\theta_j x_j} \prod_{k \in \mathcal{N}_j}\mu_{k \rightarrow j}^{(l)} (x_j)
\end{equation}
In the log domain, \eqref{mu_kjExp} and \eqref{b_jExp} convert, respectively, to 
\begin{align}
m_{k\to j}^{(l)} &= S\left(J_{kj},~ \gamma_k +\sum_{n\in {\cal N}_{k}^j}m_{n \to k}^{(l-1)}\right) \label{m_kj_2} \\
\lambda_j^{(l)} &= \gamma_j +\sum_{k\in {\cal N}_{j}}m_{k\to j}^{(l)} \label{lambda_j_2} 
\end{align}
where 
\begin{align}
\lambda_j^{(l)} &\triangleq \ln \frac{b_j^{(l)}(x_j = 1)}{b_j^{(l)}(x_j = 0)} \label{lambda_j} \\
m_{k \rightarrow j}^{(l)} &\triangleq \ln \frac{\mu_{k \rightarrow j}^{(l)} (x_j = 1)}{\mu_{k \rightarrow j}^{(l)} (x_j = 0)} \label{m_kj} 
\end{align}
denote, respectively, the estimated likelihood ratio at node $j$ and the message sent to node $j$ from node $k$ while $S(a, b)\triangleq \ln{1+e^{a+b} \over e^{a}+e^{b}}$ and $\gamma_k \triangleq \ln \frac{p(\boldsymbol{y}_k|x_k = 1)}{p(\boldsymbol{y}_k|x_k = 0)}  = \boldsymbol{s}_{k}^T \boldsymbol{y}_{k} - \frac{1}{2} \|\boldsymbol{s}_{k}\|^2$. In this model, $\boldsymbol y_k = x_k \boldsymbol s_k + \boldsymbol n_k$ denotes the signal received at node $k$. Hence, $x_k = 0$ indicates that the target signal $\boldsymbol s_k$ is absent leaving the the spectrum free where node $k$ operates. If $x_k = 1$, then the corresponding spectrum band is occupied. $J_{kj}$'s are calculated as in Eq. (16) in \cite{Abdi20} by processing a window of $T$ sensing outcomes. Note that $\theta_k$ in \eqref{m_kj_2} is merged into $\gamma_k$ without having any impact on the rest of the analysis. 

After $l^*$ iterations, $\lambda_j^{(l^*)}$ is compared, as a decision variable, against a detection threshold $\tau_j$ at node $j$ to decide the status of $x_j$, i.e., $\hat{x}_j = \boldsymbol 1\{\lambda_j^{(l^*)} - \tau_j\}$. By a linear approximation of \eqref{m_kj_2}, we have
\begin{equation}\label{m_kj_linear} 
m_{k \rightarrow j}^{(l)} \approx c_{jk} \left ( \gamma_k + \sum_{n \in \mathcal{N}_k^j}m_{n \rightarrow k}^{(l-1)}\right ) 
\end{equation}
where $c_{jk} \triangleq \frac{(e^{2J_{kj}}-1)}{(1+e^{J_{kj}})^2}$.  Consequently, we see that $\lim_{l \to \infty} \lambda_j^{(l)} \approx \lambda_j$ where  
\begin{align}\label{lambda_j_linear} 
\lambda_j  &\triangleq \gamma_j + \sum_{k \in \mathcal{N}_j}c_{jk}\gamma_k + \sum_{k \in \mathcal{N}_j}\sum_{n \in \mathcal{N}_k^j}c_{jk}c_{kn} \gamma_n \nonumber \\
&+ \sum_{k \in \mathcal{N}_j}\sum_{n \in \mathcal{N}_k^j}\sum_{m \in \mathcal{N}_n^k}c_{jk}c_{kn}c_{nm} \gamma_m + ... 
\end{align}
Therefore, we see that, given enough time, all the local likelihood ratios observed in the network are linearly combined at node $j$ to calculate its decision variable $\lambda_j$. We have shown in \cite{Abdi20} that the convergence of this linear message-passing algorithm is guaranteed when $|c_{j,k}| < \frac{1}{\operatorname*{max}_{n} |\mathcal{N}_n|-1}, \forall (j,k) \in \mathcal{E}$. The linear combination in \eqref{lambda_j_linear} can be expressed as $\lambda_j  = \sum_{i = 1}^{N} a_{ji} \gamma_i$, which is compactly stated in matrix form as
\begin{equation}\label{matrixForm} 
\boldsymbol{\lambda} = \boldsymbol{A}^T \boldsymbol{\gamma}
\end{equation}
 where $\boldsymbol{\lambda} \triangleq [\lambda_1, ..., \lambda_N]^T$ and $\boldsymbol{A} \triangleq [\boldsymbol{a}_1, ..., \boldsymbol{a}_N]$ while $\boldsymbol{a}_j \triangleq [a_{j1}, ..., a_{jN}]^T$. 
 Through some algebra, we can find the relationship between $\boldsymbol{A}$ and $c_{jk}$'s in \eqref{lambda_j_linear}. Specifically, we have
\begin{equation}\label{AandC} 
\boldsymbol{A} \approx \boldsymbol{I} + \sum_{n=1}^{\infty} \boldsymbol{C}^n - \mathfrak{D} \left (\sum_{n=1}^{\infty} \boldsymbol{C}^n \right )
\end{equation}
where $\boldsymbol{C} \triangleq \left [ c_{jk} \right ]_{N \times N}$ and $\mathfrak{D}(\boldsymbol{X})$ denotes a diagonal matrix whose main diagonal is equal to that of $\boldsymbol{X}$. The proof is provided in Appendix I. 

It is now clear that to have convergence in the message-passing iteration \eqref{m_kj_linear}, the spectral radius of $\boldsymbol{C}$ has to be less than one. This criterion may be used to impose bounds on $c_{jk}$'s to guarantee the convergence of the algorithm. Alternatively, the convergence can be guaranteed, without dealing with the complexities of finding the spectral radius, by using the contracting mapping condition as we have discussed in \cite{Abdi20}. We use \eqref{AandC} in the following section to derive an estimation of the error strength affecting the decision variables built by an erroneous BP.

\section{Errors in Belief Propagation} \label{sec:errorsInBP}
Eq. \eqref{m_kj_2} shows that at each BP iteration each node creates its messages in terms of its local LLR value as well as the messages received from the neighboring nodes at the previous iteration. In our system model, we assume that the local LLRs and the BP messages are erroneous. As in \cite{Huang15, Ihler05}, we use the von Neumann approach to modeling the joint statistical behavior of errors.

\subsection{Error Model and Analysis}
 Since the messages are multiplied together to build the beliefs, we formulate them as multiplicative perturbations affecting true (i.e., error-free) message values, i.e.,  
\begin{equation}\label{eq:mu_kj_err} 
\tilde{\mu}_{k \to j}^{(l)}(x_j) = \mu_{k \to j}^{(l)}(x_j)\varepsilon_{k \to j}^{(l)}(x_j)
\end{equation}
where $\tilde{\mu}_{k \to j}^{(l)}(x_j)$ denotes the erroneous message sent to node $j$ from node $k$ at iteration $l$ while $\varepsilon_{k \to j}^{(l)}(x_j)$ denotes the corresponding error, which is considered in this paper as a stochastic process. 

\emph{Remark 1}: Eq. \eqref{eq:mu_kj_err} differs from the model used in \cite{Ihler05} in the sense that the error model in that work measures the difference between the messages at iteration $l$ with their counterparts at the fixed point of the message-passing iteration. In other words, the error model in \cite{Ihler05} measures the deviation of the messages at each iteration from their final value reached by BP after convergence. The stochastic error we discuss here is briefly studied in \cite{Ihler05} under the notion of \emph{additional error}. 

 By expressing the messages in the the log domain we have 
\begin{equation}\label{eq:m_kj_err} 
\tilde{m}_{k \to j}^{(l)} \triangleq \ln \frac{\tilde{\mu}_{k \to j}^{(l)} (x_j = 1)}{\tilde{\mu}_{k \to j}^{(l)} (x_j = 0)} = m_{k \to j}^{(l)} + \nu_{k \to j}^{(l)}
\end{equation}
where 
\begin{equation}\label{eq:nu_kj} 
\nu_{k \to j}^{(l)} \triangleq \ln \frac{\varepsilon_{k \to j}^{(l)} (x_j = 1)}{\varepsilon_{k \to j}^{(l)} (x_j = 0)}
\end{equation}
Based on the von Neumann model, we assume that if $k \neq n$, then $E[\ln \varepsilon_{k \to j}^{(l)}(x) \ln \varepsilon_{n \to j}^{(l)}(x)] = 0$ for all $x$. Consequently, we have $E[\nu_{k \to j}\nu_{n \to j}] = 0$. To measure the collective impact of errors on the belief of node $j$, we use 
\begin{equation} \label{eq:E_j} 
E_j^{(l)}(x_j) \triangleq {\tilde{b}_j^{(l)}(x_j) \over b_j^{(*)}(x_j)}
\end{equation}
where $\tilde{b}_j^{(l)}(x_j)$ denotes the belief at node $j$ resulting from a BP iteration with erroneous messages as in \eqref{eq:mu_kj_err} while $b_j^{(*)}(x_j)$ denotes the belief of node $j$ at a fixed point reached by an error-free BP iteration. We use $(*)$ instead of $(l)$ to indicate the messages and beliefs at a fixed point of the error-free BP. 

By assuming uncorrelated stochastic behavior for the message errors, an upper bound on cumulative errors affecting the beliefs can be obtained. Specifically, assuming $\textup{Var}\left[\nu_{k \to j}^{(l)}\right] \le (\ln u)^2$ for all $k,j,l$, an upper bound on the resulting cumulative strength of errors at node $j$ is derived in \cite{Ihler05} as,
\begin{equation} \label{eq:E_ln_d} 
\textup{E}\left [ \left\{  \ln d  \left(E_j^{(l)} \right )\right \}^2 \right ] \le \sum_{k \in \mathcal{N}_j} \left (\sigma_{kj}^{(l)}  \right )^2
\end{equation}
where $\sigma_{kj}^{(1)} = \ln d(\psi_{kj})^2$ and 
\begin{equation} \label{eq:sigma_ij} 
\left (\sigma_{kj}^{(l+1)}  \right )^2 = \left ( \ln \frac{d(\psi_{kj})^2\omega_{kj}^{(l)}+1}{d(\psi_{kj})^2 + \omega_{kj}^{(l)}} \right )^2 + (\ln u)^2
\end{equation} 
while
\begin{equation} \label{eq:xi_ij} 
\left (\ln \omega_{kj}^{(l)}  \right )^2 = \sum_{n \in \mathcal{N}_k^{j}} \left (\sigma_{nk}^{(l)}  \right )^2
\end{equation}
where 
\begin{align}
d \left (E_j^{(l)} \right ) &\triangleq \sup_{a,b} \sqrt{{E_j^{(l)}(a) \over E_j^{(l)}(b)} } \\
d(\psi_{kj})^2 & \triangleq \sup_{a,b,c,d}\frac{\psi_{kj}(a,b)}{\psi_{kj}(c,d)} 
\end{align}

We use the upper bound in \eqref{eq:E_ln_d} in the log domain based on the fact that (see \eqref{lambda_j} and  \eqref{eq:E_j})
\begin{equation} \label{eq:xi_ij} 
\tilde{\lambda}_j^{(l)} \triangleq \ln \frac{\tilde{b}_j^{(l)}(1)}{\tilde{b}_j^{(l)}(0)} = \lambda_j^{(*)}+ \ln {E_j^{(l)}(1) \over E_j^{(l)}(0)} 
\end{equation}
which leads to 
\begin{align} \label{eq:ihler'sBound} 
\textup{E} \left [\left |\tilde{\lambda}_j^{(l)} - \lambda_j^{(*)} \right|^2  \right ] &= \textup{E} \left [ \left| \ln E_j^{(l)}(1) - \ln E_j^{(l)}(0) \right|^2 \right] \nonumber \\
 &=  \textup{E}\left [ \left\{  \ln d  \left(E_j^{(l)} \right )\right \}^2 \right ] \le \sum_{k \in \mathcal{N}_j} \left (\sigma_{kj}^{(l)}  \right )^2
\end{align}
Hence, in the detection structure discussed, \eqref{eq:E_ln_d} gives an upper bound on the MSE level observed in the decision variable at node $j$. 

\subsection{Linear Approximations}
 In our analysis, we distinguish between the message errors and the errors in the computation of local LLRs to gain further insight into the behavior of the BP algorithm. In particular, we model the erroneous local LLRs as $\tilde{\gamma}_k \triangleq \gamma_k + \epsilon_k$ and refer to $\epsilon_k$'s as \emph{likelihood errors} (LE) while assuming that LEs are uncorrelated as well, i.e., $E[\epsilon_k \epsilon_n] = 0$ for $k \neq n$. We refer to $\nu_{k \to j}$'s as \emph{message errors} (ME) and assume that LEs and MEs are mutually independent. Moreover, we assume that all MEs and LEs are independent of the messages and of the local LLRs. Note that \emph{the bound in \eqref{eq:ihler'sBound} does not take LEs into account.} 

Taking both types of error into account, we express the messages as 
\begin{align} 
\tilde{m}_{k\to j}^{(l)}& = S\left(J_{kj},~ \tilde{\gamma}_k +\sum_{n\in {\cal N}_{k}^j}\tilde{m}_{n \to k}^{(l-1)}\right) + \nu_{k\to j}^{(l)}  \label{eq:m_kj_err_S} \\
\tilde{\lambda}_j^{(l)} &= \tilde{\gamma}_j +\sum_{k\in {\cal N}_{j}} \tilde{m}_{k\to j}^{(l)} \label{eq:lambda_j_err} 
\end{align}
which show that the errors pass through the same nonlinear transformation (i.e., $S$) as the messages do. By using \eqref{eq:m_kj_err_S}, we can analyze the behavior of errors. The proposed linear BP iteration in the presence of message errors is expressed as 
\begin{equation} \label{eq:m_kj_err_lin} 
\tilde{m}_{k \rightarrow j}^{(l)} \approx c_{jk} \left ( \tilde{\gamma}_k + \sum_{n \in \mathcal{N}_k^j} \tilde{m}_{n \to k}^{(l-1)}\right ) + \nu_{k \to j}^{(l)}
\end{equation}
Consequently, similar to the way \eqref{lambda_j_linear} is derived, the resulting erroneous decision variable is formed as 
\begin{equation} \label{eq:lambda_j_tilde} 
\tilde{\lambda}_j^{(l)} \approx \Big [\tilde{\gamma}_j + \sum_{k \in \mathcal{N}_j}c_{jk}\tilde{\gamma}_k + \sum_{k \in \mathcal{N}_j}\sum_{n \in \mathcal{N}_k^j}c_{jk}c_{kn} \tilde{\gamma}_n +  ...  \Big ] + \sum_{k \in \mathcal{N}_j} \nu_{k \to j}^{(l)}
\end{equation}
which can be reorganized as  
\begin{equation} \label{eq:lambda_j_tilde_short} 
\tilde{\lambda}_j^{(l)} \approx \lambda_j + \xi_j^{(l)}
\end{equation}
where 
\begin{equation}\label{eq:eta_j} 
\xi_j^{(l)} \triangleq \sum_{i=1}^{N} a_{ji}\epsilon_i + \sum_{k \in \mathcal{N}_j} \nu_{k \to j}^{(l)}
\end{equation}
Hence, we have the following remark, which we will use in Sec. \ref{sec:MitigLinFus} where we develop an optimization framework for the system. 

\emph{Remark 2}: Eq. \eqref{eq:eta_j} shows that, the error affecting the decision variable at node $j$ has two distinct components. The first component is built as a linear combination of LEs while the second one is the sum of the MEs received at node $j$ from its one-hop neighbors. The first component is fixed whereas the second one exhibits a new realization at every iteration. 

According to \eqref{eq:eta_j}, deviation from the error-free decision variables, caused by errors in the BP iterations, can approximately be measured by 
\begin{equation}\label{eq:MSE_lambda} 
 \textup{E} \left [\left |\tilde{\lambda}_j^{(l)} - \lambda_j^{(*)} \right|^2  \right ] \approx \textup{E}\left [\left |\xi_j^{(l)}\right|^2 \right ] = \boldsymbol{a}_j^T \boldsymbol{\Sigma}_{\boldsymbol{\epsilon}} \boldsymbol{a}_j + \textup{tr} \left (\boldsymbol{\Sigma}_{\boldsymbol{\nu}_j}  \right )
\end{equation}
where $\boldsymbol{\Sigma}_{\boldsymbol{\epsilon}} \triangleq \textup{cov}(\boldsymbol{\epsilon})$ and $\boldsymbol{\Sigma}_{\boldsymbol{\nu}_j} \triangleq \textup{cov} \left (\boldsymbol{\nu}_j^{(l)} \right )$ while $\boldsymbol{\epsilon} \triangleq [\epsilon_1, ..., \epsilon_N]^T $ and $\boldsymbol{\nu}_j^{(l)}$ denotes an $|\mathcal{M}_j|$-by-1 vector that contains $\nu_{k \to j}^{(l)}$'s for $k \in \mathcal{M}_j$ where ${\cal M}_{j} \triangleq  \mathcal{N}_j \cup \{j\}$ while $\nu_{j \to j}^{(l)} \triangleq 0$. 

\newcommand{\Conv}{%
  \mathop{\scalebox{1.5}{\raisebox{-0.2ex}{$\circledast$}}
  }
}

Eq. \eqref{eq:lambda_j_tilde} shows that when BP is used to realize a distributed detection, the erroneous local likelihoods in the network are combined linearly to build the decision variables. We can evaluate the impact of the errors on the system performance by analyzing the stochastic behavior of the erroneous decision variables $\tilde{\lambda}_j^{(l)}$. Given $\boldsymbol{x}$, the decision variable at node $j$ is obtained as a linear combination of independent random variables. Consequently, its conditional pdf is derived as
\begin{equation} \label{eq:f_lambda} 
f_{\tilde{\lambda}_j | \boldsymbol{x}}(z|\boldsymbol{b}) \approx \Conv_{i = 1}^{N}{1 \over a_{ji}}f_{\tilde{\gamma}_i| \boldsymbol{x}}\left({z \over a_{ji}} \vert\boldsymbol{b} \right) *  \Conv_{k \in \mathcal{N}_j}f_{\nu_{k \to j}}(z)
\end{equation}
where 
\begin{equation} \label{eq:f_gamma} 
f_{\tilde{\gamma}_i|\boldsymbol{x}}(z|\boldsymbol{b}) = f_{\gamma_i|\boldsymbol{x}}(z|\boldsymbol{b})*f_{\epsilon_i}(z)
\end{equation}
 while $ \Conv$ and $*$ denote the convolution operator. Consequently, we have 
\begin{align}  \label{eq:g_j}
\textsl{g}_j(\tau_j,v) &\triangleq \textup{Pr}\{\tilde{\lambda}_j > \tau_j | x_j = v\}   \nonumber \\
&=\sum_{\boldsymbol{b} \in \{0,1\}^{N-1}} p_{\boldsymbol{x}_{(j)} |x_j}
\left(\boldsymbol{b}|v \right)  \int_{\tau_j}^\infty f_{\tilde{\lambda}_j | \boldsymbol{x}} \left (z|\mathcal{E}_{j,v}(\boldsymbol{b})\right)dz 
\end{align}
where $v \in \{0,1\}$, $\boldsymbol{x}_{(j)} \triangleq [x_1, x_2, ..., x_{j-1},  x_{j+1}, ..., x_N]^T$ and $\mathcal{E}_{j,v}(\boldsymbol{b}) \triangleq \{\boldsymbol{x}_{(j)}=\boldsymbol{b}, x_j = v\}$ while $p_{\boldsymbol{x}_{(j)}|x_j}(\boldsymbol{b}|v) \triangleq \textup{Pr}\{\boldsymbol{x}_{(j)} = \boldsymbol{b} \vert x_j = v\}$. Solving $\textsl{g}_j(\tau_j,0) = \alpha$ gives a threshold value that fixes the false-alarm rate at $\alpha$. Similarly, $\textsl{g}_j(\tau_j,1) = \beta$ fixes the detection rate at $\beta$. Recall that $a_{ji}$'s are found by using $c_{jk}$'s, see \eqref{AandC}.

As a common practical case, when the local LLRs and the errors follow Gaussian distributions \cite{Quan08, Quan09, Quan10, Taricco11, Abdi14} the decision variable $\tilde{\lambda}_j$ follows a Gaussian distribution as well and it is fully characterized by its first- and second-order statistics. Specifically, we have 
\begin{equation}  \label{eq:integ_f}
  \int_{\tau_j}^\infty f_{\tilde{\lambda}_j | \boldsymbol{x}} \left (z|\mathcal{E}_{j,v}(\boldsymbol{b})\right)dz 
 = Q\left ( \frac{\tau_j - \mu_{j,v}(\boldsymbol{b}) }{\sigma_{j,v}(\boldsymbol{b}) } \right )
\end{equation}
where 
\begin{align}
\mu_{j,v}(\boldsymbol{b})  &\triangleq  \textup{E}\left[\tilde{\lambda}_j |\mathcal{E}_{j,v}(\boldsymbol{b}) \right] 
\nonumber \\
&= \textup{E}\left[\gamma_j|x_j = v\right] + \sum_{ i\neq j} a_{ji}\textup{E}\left[\gamma_i|x_i = b_i\right]  \label{eq:mu(b)} \\
\sigma_{j,v}^2(\boldsymbol{b}) &\triangleq \textup{Var}\left[\tilde{\lambda}_j |\mathcal{E}_{j,v}(\boldsymbol{b}) \right] \nonumber \\
&= \textup{Var}\left[\gamma_j|x_j = v\right] + \sum_{i \neq j} a_{ji}^2 \textup{Var}\left[\gamma_i|x_i = b_i\right]+ \textup{E}\left [|\xi_j|^2 \right ] 
 \label{eq:sigma(b)}
\end{align}
In \eqref{eq:mu(b)} we have assumed, without loss of generality, to have zero-mean errors. Note that, without the proposed approximation these performance measures are not available analytically due to the nonlinearity of \eqref{m_kj_2}. In the rest of the paper, we assume that the local likelihoods, LEs, and MEs are  Gaussian random variables. Eq. \eqref{eq:f_lambda} shows that, according to the central limit theorem \cite{Papoulis}, even if the local LLRs and errors are not Gaussian random variables, the stochastic behavior of the decision variables can still be approximately described by Gaussian distributions.

\subsection{Impact of Averaging} \label{subsec:Ave}
 In ABP, the message-passing iteration is the same as in BP. However, instead of the actual message values, an average of the messages are used to build the decision variables. To be more specific, in the log domain and for $l \ge L+1$, let
\begin{equation}\label{eq:m_kj_bar} 
\bar{m}_{k \to j}^{(l)} \triangleq {1 \over {L+1}} \sum_{t=l-L}^{l} \tilde{m}_{k \to j}^{(t)}
\end{equation}
The decision variable at node $j$ is calculated by 
\begin{equation} \label{eq:lambda_j_bar} 
\bar{\lambda}_j^{(l)} \triangleq \gamma_j + \sum_{k \in \mathcal{N}_j}\bar{m}_{k \to j}^{(l)}
\end{equation}
Similar to our discussion regarding \eqref{lambda_j_linear}, we can show that when the message-passing iteration is error-free,  $\bar{\lambda}_j^{(*)} \triangleq \lim_{l \to \infty} \bar{\lambda}_j^{(l)} = \lambda_j$. Hence, we have the following remark. 

\emph{Remark 3}: The averaging process does not alter the fixed points achieved by the error-free linear BP. From an approximation-based point of view, this observation is in line with the convergence analysis provided in \cite{Huang15}. 

The impact of averaging on LEs and MEs can be clarified by noting that 
\begin{equation} \label{eq:lambda_j_bar_short} 
\bar{\lambda}_j^{(l)} = \lambda_j + \bar{\xi}_j^{(l)}
\end{equation}
where, assuming $L$ to be large enough, we have
\begin{equation} \label{eq:eta_j_bar} 
\bar{\xi}_j^{(l)} = \sum_{i=1}^{N} a_{ji}\epsilon_i + \sum_{k \in \mathcal{N}_j} \bar{\nu}_{k \to j}^{(l)} \approx \sum_{i=1}^{N} a_{ji}\epsilon_i 
\end{equation}
since $\bar{\nu}_{k \to j}^{(l)} \triangleq {1 \over {L+1}} \sum_{t=l-L}^{l} \nu_{k \to j}^{(t)} \approx 0 $. We can state \eqref{eq:eta_j_bar} in the form of MSE as 
\begin{equation}\label{eq:MSE_lambda} 
 \textup{E} \left [\left |\bar{\lambda}_j^{(l)} - \lambda_j^{(*)} \right|^2  \right ] \approx \boldsymbol{a}_j^T \boldsymbol{\Sigma}_{\boldsymbol{\epsilon}} \boldsymbol{a}_j + {1 \over L+1}\textup{tr} \left (\boldsymbol{\Sigma}_{\boldsymbol{\nu}_j}  \right )
\end{equation}

\emph{Remark 4}: Assuming $L$ to be large enough and the MEs to have zero mean, \eqref{eq:eta_j_bar} shows that the resulting decision variable built by ABP in \eqref{eq:lambda_j_bar} is almost cleared of MEs. However, the averaging process has almost no impact on LEs. 

Note that in ABP the message-passing iteration is the same as in BP and the averaging is only performed when computing the decision variables. Moreover, in ABP, instead of storing the messages in past iterations separately, we only need to store the sum of the messages up to the current iteration. As a consequence, the number of additional memory cells required can be kept constant \cite{Huang15}. We will use ABP in Sec. \ref{subsec:learnOptim} to build an offline learning-optimization structure for the linear BP in the presence of errors. 

\section{Mitigating Errors by Linear Fusion} \label{sec:MitigLinFus}
In this section, we first propose a two-stage linear fusion scheme to obtain a near-optimal detection performance by suppressing the impact of the errors. Then, we realize the proposed optimization in a blind decentralized setting where the required statistics are not available a priori.

\subsection{Linear Fusion}
  First, since $|c_{jk}| < 1$, we further approximate the decision variable $\lambda_j$ in \eqref{lambda_j_linear}  as 
 \begin{equation} \label{eq:lambda_j_approx} 
\lambda_j \approx  \sum_{k\in {\cal M}_{j}}c_{jk}\gamma_k 
\end{equation} 
 Due to the symmetry of the data-fusion process in \eqref{lambda_j_linear}, the approximation in \eqref{eq:lambda_j_approx} is an effective approach to building a distributed computing framework for the system performance optimization. In this framework, each node interacts only with its immediate neighbors. We have clarified this symmetry in \cite[Sec. III-B]{Abdi20}. By taking into account the errors while analyzing the linear BP, \eqref{lambda_j_linear} and \eqref{eq:lambda_j_tilde} lead to  
 \begin{equation} \label{eq:lambda_j_tilde_approx} 
\tilde{\lambda}_j \approx  \sum_{k\in {\cal M}_{j}}c_{jk}\left (\gamma_k + \epsilon_k  \right ) + \sum_{k\in {\cal N}_{j}}\nu_{k \to j}
\end{equation} 

We see that the disturbance on the decision variable caused by LEs is built, approximately, as a linear combination of $\epsilon_k$'s with $c_{jk}$'s acting as weights in this combination. Therefore, we use  $c_{jk}$'s as design parameters to mitigate the impact of $\epsilon_k$'s. Moreover, MEs are combined in \eqref{eq:lambda_j_tilde_approx} linearly and in this combination, all weights are one. We propose to extend this combination by using a modified version of \eqref{eq:lambda_j_err}  as 
 \begin{equation} \label{eq:lambda_j_hat} 
\hat{\lambda}_j^{(l)} \triangleq \tilde{\gamma}_j +\sum_{k\in {\cal N}_{j}} w_{jk}\tilde{m}_{k\to j}^{(l)} 
\end{equation} 
This modification in the structure of the decision variable does not affect the convergence of the proposed linear BP since it does not alter the message-passing iteration. Now, based on an approximation similar to the one in \eqref{eq:lambda_j_tilde_approx}, we have 
\begin{equation} \label{eq:lambda_j_hat_long} 
\hat{\lambda}_j \approx  \sum_{k\in {\cal M}_{j}}w_{jk}c_{jk}\left (\gamma_k + \epsilon_k  \right ) + \sum_{k\in {\cal N}_{j}} w_{jk}\nu_{k \to j}
\end{equation} 

Since $\hat{\lambda}_j$ is a Gaussian random variable, we only need its mean and variance to characterize its statistical behavior. Specifically, for $b \in \{0,1\}$, we have 
\begin{equation} \label{eq:Pr_lambda} 
\textup{Pr}\{\hat{\lambda}_j > \tau_j|x_j = b\} = Q \left (\frac{\tau_j - \textup{E}[\hat{\lambda}_j|x_j = b]}{\textup{Var}[\hat{\lambda}_j|x_j = b]}  \right )
\end{equation} 
where
\begin{align}
\textup{E}[\hat{\lambda}_j|x_j = b] & \approx \boldsymbol{v}_j^T \boldsymbol{\mu}_b \label{eq:E[lambda]} \\
\textup{Var}[\hat{\lambda}_j|x_j = b] & \approx \boldsymbol{v}_j^T \left (\boldsymbol{\Sigma}_{\boldsymbol{\gamma_j}|b} + \boldsymbol{\Sigma}_{\boldsymbol{\epsilon_j}}  \right )   \boldsymbol{v}_j +\boldsymbol{w}_j^T \boldsymbol{\Sigma}_{\boldsymbol{\nu_j}} \boldsymbol{w}_j \label{eq:Var[lambda]}
\end{align}
where $\boldsymbol{v}_j \triangleq \boldsymbol{w}_j \circ \boldsymbol{c}_j$ in which $\circ$ denotes the Hadamard product while $\boldsymbol{\Sigma}_{\boldsymbol{\gamma}_j|b} = \textup{cov}(\boldsymbol{\gamma}_j|x_j = b)$ and $\boldsymbol{\Sigma}_{\boldsymbol{\epsilon}_j} = \textup{cov}(\boldsymbol{\epsilon}_j) $.  Moreover, $\boldsymbol{w}_j$, $\boldsymbol{c}_j$, $\boldsymbol{\gamma}_j$, and $\boldsymbol{\epsilon}_j$ are $|\mathcal{M}_j|$-by-1 vectors containing $w_{ji}$'s, $c_{ji}$'s, $\gamma_i$'s, and $\epsilon_i$'s for $i \in \mathcal{M}_j$, respectively. Eq. \eqref{eq:Pr_lambda} gives the system false-alarm probability for $b = 0$ and the detection probability for $b = 1$. The false-alarm probability can be set to $P_{\textup{f}}^{(j)} = \alpha$ by
\begin{equation}\label{eq:tau_j}
\tau_j = Q^{-1}(\alpha) \textup{Var}[\hat{\lambda}_j|x_j = 0] + \textup{E}[\hat{\lambda}_j|x_j = 0]
\end{equation}
and then by using \eqref{eq:Pr_lambda} -- \eqref{eq:Var[lambda]},  $\boldsymbol{w}_j $ and $\boldsymbol{c}_j $ can jointly be optimized in a Neyman-Pearson setting. 

 In order to avoid the challenges associated with this optimization, we  maximize the deflection coefficient of the detector. We already know that the resulting detector performs well when the decision variables follow the Gaussian distribution. In this manner, we mitigate the joint impact of LEs and MEs with low computational complexity. 
 
 The proposed optimization is conducted in two consecutive stages based on the fact that we can decompose the construction of $\hat{\lambda}_j$ into two consecutive fusion processes. That is, we first optimize $c_{jk}$'s by considering the impact of $\epsilon_k$'s on $\gamma_k$'s. Then, we consider the resulting scaled LLRs, i.e.,  $c_{jk} \gamma_k$'s, as new statistics to be linearly combined, while being weighted by $w_{jk}$'s and distorted by $\nu_{k \to j}$'s, to make the decision variable at node $j$. 
 
 More specifically, first, we optimize $\boldsymbol{c}_j$ in a hypothetical linear detector with its decision variable defined as 
  \begin{equation} \label{eq:lambda'} 
\hat{\lambda}_j' \triangleq \boldsymbol{c}_j^T \left (\boldsymbol{\gamma}_j + \boldsymbol{\epsilon}_j  \right ) 
\end{equation} 
 The coefficients resulting from this optimization scale up the more reliable local LLRs, with respect to the ones built under low SNR regimes, to suppress the effect of LEs. We denote the resulting fusion weights by $\boldsymbol{c}_j^*$. Then, we use $\boldsymbol{c}_j^*$ within the structure of the actual detector to optimize $\boldsymbol{w}_j$ to mitigate the impact of MEs. That is, we consider the following linear detector at node $j$ 
 \begin{equation} \label{eq:lambda''} 
\hat{\lambda}_j'' \triangleq \boldsymbol{w}_j^T\left (\boldsymbol{\chi}_j + \boldsymbol{\nu}_j  \right )
\end{equation} 
where $\boldsymbol{\chi}_j \triangleq \boldsymbol{c}_j^* \circ (\boldsymbol{\gamma}_j + \boldsymbol{\epsilon}_j)$ contains $\chi_{jk}$'s for $k \in \mathcal{M}_j$ while $\chi_{jk} = c_{jk}^*(\gamma_k + \epsilon_k)$. The vector $\boldsymbol{\nu}_j$ contains $\nu_{k \to j}$'s with $k \in \mathcal{M}_j$. 
In this structure, the elements of $\chi_{jk}$'s are seen as the actual local LLRs that are combined to build the decision variable at node $j$ while the combination takes into account the joint degrading effect of MEs and LEs. 

Based on the material provided in Sec. \ref{sec:linearFusion}, the first stage of the proposed optimization is formally stated as
\begin{equation} \label{eq:c_opt} 
\begin{matrix}
 \boldsymbol{c}_j^* = \arg \max_{\boldsymbol{c}_j} \Delta_j'(\boldsymbol{c}_j)& \textup{s.t.,}~ \left \| \boldsymbol{c}_j \right \| = 1  
\end{matrix}
\end{equation} 
where 
\begin{equation} \label{eq:Delta'} 
\Delta_j'(\boldsymbol{c}_j) = \frac{\left (\boldsymbol{c}_j^T \boldsymbol{\delta}_j  \right )^2 }{\boldsymbol{c}_j^T \left (\boldsymbol{\Sigma}_{\boldsymbol{\gamma}_j|0} + \boldsymbol{\Sigma}_{\boldsymbol{\epsilon}_j}  \right )~ \boldsymbol{c}_j }
\end{equation} 
where $\boldsymbol{\delta}_j \triangleq \textup{E}[\boldsymbol{\gamma}_j|x_j = 1] - \textup{E}[\boldsymbol{\gamma}_j|x_j = 0]$. 
The resulting $\boldsymbol{c}_j^*$ is then used to realize the second stage of the proposed optimization by solving 
\begin{equation} \label{eq:w_opt''} 
\begin{matrix}
 \boldsymbol{w}_j^* = \arg \max_{\boldsymbol{w}_j} \Delta_j''(\boldsymbol{w}_j)& \textup{s.t.,}~ \left \| \boldsymbol{w}_j \right \| = 1
\end{matrix}
\end{equation} 
where 
\begin{equation} \label{eq:Delta''} 
\Delta_j'' (\boldsymbol{w}_j) = \frac{\left (\boldsymbol{w}_j^T \hat{\boldsymbol{\delta}}_j   \right )^2 }{\boldsymbol{w}_j^T \left (\boldsymbol{\Sigma}_{\boldsymbol{\chi}_j|0} + \boldsymbol{\Sigma}_{\boldsymbol{\nu}_j}  \right )~ \boldsymbol{w}_j }
\end{equation}
where $\hat{\boldsymbol{\delta}}_j = \boldsymbol{c}_j^* \circ \boldsymbol{\delta}_j$ and $\boldsymbol{\Sigma}_{\boldsymbol{\chi}_j|0} = \textup{cov}(\boldsymbol{\chi}_j|x_j = 0) =  \boldsymbol{c}_j^* \boldsymbol{c}_j^{*T} \circ \left (\boldsymbol{\Sigma}_{\boldsymbol{\gamma}_j|0} + \boldsymbol{\Sigma}_{\boldsymbol{\epsilon}_j} \right )$. Having $\boldsymbol{c}_j^*$ and $\boldsymbol{w}_j^*$, the detection threshold $\tau_j$ is derived as $\tau_j = Q^{-1}(\alpha) \textup{Var}[\lambda_j''|x_j = 0] + \textup{E}[\lambda_j''|x_j = 0]$ to fix the system false-alarm rate at $\alpha$.

\emph{Remark 5}: The convergence condition $|c_{j,k}| < \frac{1}{\operatorname*{max}_{n} |\mathcal{N}_n|-1}, \forall (j,k) \in \mathcal{E}$ can be realized by a simple normalization of $c_{j,k}^*$'s since the objective function in  \eqref{eq:c_opt} does not change by normalizing its argument.

Through the proposed two-stage optimization, we enhance the detection performance at node $j$ by suppressing the joint impact of MEs and LEs with low computational complexity. The statistics required in this optimization are collected from the one-hop neighbors of node $j$. This makes the proposed method a viable approach in ad-hoc network configurations where major network functionalities are conducted through one-hop links between the network nodes.  

\subsection{Offline Learning and Adaptation} \label{subsec:learnOptim}
To realize the proposed optimization, we need the mean and covariance of the local erroneous LLRs. In a blind setting where there is no prior information available regarding the radio environment, we have to estimate those parameters based on the detection outcomes. In particular, the main challenge here is that the state of $x_j$ is required at node $j$ while the only information available in practice is the detection outcome $\hat x_j$. Hence, node $j$ has to estimate the conditional statistics required in \eqref{eq:c_opt} and \eqref{eq:w_opt''} based on $\hat x_j$. The problem with such an adaptation mechanism is that it makes the detection outcome $\hat x_j$ depend on those estimates. This dependence creates an inherent deteriorating loop by feeding the detection errors back into the system structure through erroneous estimates of the required statistics. 

To overcome this challenge, we propose an extended version of the blind learning-adaptation loop in \cite{Abdi20} that accommodates the proposed error-mitigating structure. The pseudo-code of this adaptation is provided in Algorithm \ref{tab:offlineBP} where the task of each node is specified in a distributed computing framework. Algorithm \ref{tab:offlineBP} operates on a window of stored sensing outcomes and involves a secondary BP that is run much less frequently than the rate at which the distributed detection is performed. The outcomes of this offline BP are used in the estimation of the required unknown statistics. In this adaptation, the desired optimizations are realized iteratively while each node interacts only with its one-hop neighbors. Consequently, Algorithm \ref{tab:offlineBP} can be well incorporated in a decentralized network configuration. 

\captionsetup[table]{name= Algorithm}
\begin{table}[]
\centering
\caption{Blind Adaptation of Fusion Weights in Erroneous Linear Belief Propagation}
\label{tab:offlineBP}
\begin{tabular}{|l|l|l|l|l|}
\hline
Input: $\tilde{\boldsymbol{\gamma}}_{T}$, $\bar{\boldsymbol{\gamma}}_{T}$,$\boldsymbol{\tau}^{(0)}$, $\kappa_{\textup{max}}$, $\eta$ \\ 
Output: Near-optimal $\boldsymbol{c}_j$ and $\boldsymbol{w}_j$ for $j = 1, ..., N$
 \\ \thickhline 
 
1. \hspace{1mm} Let $\kappa \gets 0$ and initialize $\hat{\boldsymbol{x}}^{(0)}$ by comparing $\tilde{\boldsymbol{\gamma}}_{T}$ against $\boldsymbol{\tau}^{(0)}$ ; \\

2. \hspace{1mm} \textbf{while} $\kappa \le \kappa_{\textup{max}}$ \\

3. \hspace{3mm} \textbf{for} node $j \in \{1,2, ..., N\}$ \\

4. \hspace{5mm} Calculate $E[\bar{\gamma}_i \vert \hat{x}_j^{(\kappa)}]$ and $\textup{cov}(\bar{\gamma}_i, \bar{\gamma}_k  \vert \hat{x}_j^{(\kappa)})$ for all $i,k \in \mathcal{M}_j$;  \\

5. \hspace{5mm} Solve \eqref{eq:c_opt} to find $\boldsymbol{c}_j^{(\kappa)}$ and  $\tau_j^{(\kappa)}$;\\

 6. \hspace{5mm}  Set $\boldsymbol{c}_j^*$ by an $\eta$-test on $\boldsymbol{c}_j^{(\kappa)}$; \\

 7. \hspace{3mm}  \textbf{end}\\
  
8. \hspace{3mm} Use $\boldsymbol{c}_j^*$'s and $\tau_j^{(\kappa)}$'s to run linear ABP on $\tilde{\boldsymbol{\gamma}}_{T}$ to find $\hat{\boldsymbol{x}}^{(\kappa+1)}$; \\

9. \hspace{3mm}  $\kappa \gets \kappa + 1$;\\
 
10. \hspace{1mm} \textbf{end}\\
 
11. \hspace{3mm} \textbf{for} node $j \in \{1,2, ..., N\}$ \\
 
12. \hspace{5mm} Use $\boldsymbol{c}_j^*$ and $\hat{\boldsymbol{x}}^{(\kappa_{\textup{max}})}$ to calculate $\hat{\boldsymbol{\delta}}_j$, $\boldsymbol{\Sigma}_{\boldsymbol{\chi}_j|0}$ and $\boldsymbol{\Sigma}_{\boldsymbol{\nu}_j}$; \\

% Find $E[\bar{\gamma}_i \vert \hat{x}_j^{(\kappa_{\textup{max}})}]$ and $\textup{cov}(\bar{\gamma}_i, \bar{\gamma}_k  \vert \hat{x}_j^{(\kappa_{\textup{max}})})$ for all $i,k \in \mathcal{M}_j$;  \\

 13. \hspace{5mm} Solve \eqref{eq:w_opt''}  to find $\boldsymbol{w}_j^*$;\\
 
 14. \hspace{1mm} \textbf{end}\\ 

15. \hspace{1mm} Output $\boldsymbol{c}_j^*$ and $\boldsymbol{w}_j^*$ for $j \in 1,2,...N$;

%\\\hline
%This is another test. 
\\\hline
\end{tabular}
\end{table}

In the sequel, we propose a blind adaptation structure in which we use $\kappa$ to denote the iteration index. Note that we use $l$ as the iteration index in the main BP through which the distributed detection is realized. The offline adaptation updates the fusion weights in the proposed linear BP by processing $T$ stored samples of $\tilde{\boldsymbol{\gamma}}$. This window of erroneous local likelihoods is denoted by $\tilde{\boldsymbol{\gamma}}_{T}$ and contains samples of $\tilde{\boldsymbol{\gamma}}(t)$ for $t = 1,2,..., T$. Recall that, $\tilde{\boldsymbol{\gamma}} = \boldsymbol{\gamma} + \boldsymbol{\epsilon}$ where $\boldsymbol{\epsilon}$ denotes the vector of LEs.  The offline detection outcomes at iteration $\kappa$ are denoted by $\hat{\boldsymbol{x}}^{(\kappa)} \triangleq [\hat{x}_1^{(\kappa)}, ..., \hat{x}_N^{(\kappa)}]$ while the resulting fusion weights and detection thresholds are denoted $\boldsymbol{c}_j^{(\kappa)}$ and $\tau_j^{(\kappa)}$ respectively. $\hat{\boldsymbol{x}}^{(\kappa)}$ denotes a window of stored sensing outcomes $\hat{\boldsymbol{x}}^{(\kappa)}(t)$ for $t = 1,2,..., T$.  For simplicity, we do not show the time index when dealing with $\tilde{\boldsymbol{\gamma}}_{T}$, and $\hat{\boldsymbol{x}}^{(\kappa)}$. 

Due to errors caused by the wireless links between the sensing nodes, node $j$ does not have access to $\tilde\gamma_k(t)$, $k \in \mathcal N_j$. Specifically, what node $j$ receives from node $k$ is $\tilde\gamma_k(t) + \nu_{k \to j}$ where $\nu_{k \to j}$ denotes the corresponding link error. Without loss of generality, we attribute MEs to wireless link errors. To alleviate the link errors, before starting the adaptation process node $j$ receives $L$ copies of $\tilde\gamma_k(t)$ from node $k$ and calculates an average to obtain $\bar\gamma_k(t) \triangleq \tilde\gamma_k(t) + \bar\nu_{k \to j}$ where $\bar\nu_{k \to j}$ denotes the average of $L$ independent realizations of $\nu_{k \to j}$. The desired statistics are then calculated by processing $\bar\gamma_k$'s, which approximate $\tilde\gamma_k$'s. We use $\bar{\boldsymbol{\gamma}}_T$ to contain the samples of $\bar\gamma_k(t)$ for $t = 1,2, ..., T$ for $k = 1,2,...,N$. 

In a realistic detection scenario, the data exchanged between the nodes in the proposed offline adaptation is impaired by both types of errors. Since in the first linear fusion \eqref{eq:c_opt} we take into account the impact of LEs only, we need to isolate this optimization from the MEs. To this end, we estimate the desired statistics by using linear ABP. As we saw in Sec. \ref{subsec:Ave}, MEs do not affect the ABP outcomes significantly. Therefore, the resulting offline decision variables are almost cleared of MEs and closely realize \eqref{eq:c_opt}. Note that, the offline linear ABP processes $\tilde{\boldsymbol{\gamma}}_{T}$ (not $\bar{\boldsymbol{\gamma}}_T$) since each node, say node $j$, builds its own messages by using its own local likelihood $\tilde\gamma_j$. The outcomes of the linear ABP are then used in processing $\bar{\boldsymbol{\gamma}}_T$ as indicated in line 4 of Algorithm \ref{tab:offlineBP}. 

As indicated in line 5, based on the outcomes of the ABP, we suppress the impact of LEs by the first-stage linear fusion. The resulting fusion coefficients enhance the quality of the linear ABP, which, in turn, enhances the quality of the fusion coefficients obtained in the following iteration. By repeating this learning-optimization cycle, we suppress the impact of LEs significantly while this cancellation process is not disturbed by the MEs. See lines 2 -- 10 in Algorithm \ref{tab:offlineBP}. 

In this section, we distinguish between the coefficients obtained by Algorithm I and the ones obtained by linearzing the BP algorithm as in \eqref{m_kj_linear}. Specifically, we use  $\mathbf{c}_j^{\textup{BP}} $  to collect $c_{jk}^{\textup{BP}}$'s for $k \in \mathcal{M}_j$ while $c_{jk}^{\textup{BP}} \triangleq (e^{2J_{kj}}-1)/(1+e^{J_{kj}})^2$. Line 6 indicates what we refer to as the $\eta$-\emph{test} that ensures the system performance level not to fall below that of the legacy BP algorithm. The test is as follows: given a predefined value $\eta$ and for $n = 1,2, ...,N$, if $\boldsymbol c_j^{\textup{BP}}(n)/\boldsymbol c_j^{(\kappa)}(n) \ge \eta$ then $\boldsymbol c_j^*(n) \gets \boldsymbol c_j^{\textup{BP}}(n)$. Otherwise, $\boldsymbol c_j^*(n) \gets \boldsymbol c_j^{(\kappa)}(n)$. That is, we do not use a coefficient obtained by the offline optimization if that coefficient is not large enough with respect to its corresponding coefficient in the main linearized BP. The reason is that, when the primary-user signal received at node $j$ is buried under heavy noise, the local optimization at node $j$ is not able to fully capture the correlations between $x_j$ and its neighbors. Consequently, the resulting $c_{jk}$'s attain values too small to maintain a good detection performance. In such cases, we replace those coefficients with their counterparts in the linearized BP. This technique prevents the resulting coefficients from degrading the performance when the corresponding nodes operate under an SNR regime that is not good enough for a reliable estimation of the desired statistics. We use the legacy BP coefficients for those nodes. 

Having the high-quality fusion weights $\boldsymbol{c}_j^*$ and more-reliable detection outcomes $\hat{\boldsymbol{x}}^{(\kappa_{\textup{max}})}$, we realize the second stage of the proposed optimization. To this end, node $j$ finds $\textup{cov}(\boldsymbol{\chi}_j|x_j = 0)$ in \eqref{eq:Delta''} by calculating $\textup{cov}(\bar{\boldsymbol \gamma}_j \vert \hat x_j^{(\kappa_{\textup{max}})}=0)$ and then multiplying the result, element-wise, by $\boldsymbol{c}_j^* \boldsymbol{c}_j^{*T}$. Note that $\textup{cov}(\bar{\boldsymbol \gamma}_j \vert x_j) \approx \boldsymbol{\Sigma}_{\boldsymbol{\gamma}_j|x_j} + \boldsymbol{\Sigma}_{\boldsymbol{\epsilon}_j}$ where $\bar{\boldsymbol \gamma}_j$ denotes the $j$th column in $\bar{\boldsymbol \gamma}_T$. The elements of the diagonal matrix $\boldsymbol{\Sigma}_{\boldsymbol{\nu}_j}$ are found at node $j$ by noting that $\bar\gamma_k(t) \approx \tilde\gamma_k(t)$, which indicates that $\textup{Var} [\nu_{k \to j}] \approx \textup{Var} [\tilde\gamma_k(t) + \nu_{k \to j}] - \textup{Var} [\bar\gamma_k(t)]$. Lines 11 -- 14 in Algorithm \ref{tab:offlineBP} indicate the second stage of the proposed optimization. 

In case a certain performance level, such as a certain false-alarm rate, is required to be guaranteed, we can use the  \emph{detector calibration} technique in \cite{Abdi20}. The thresholds obtained by Algorithm I appear to be  too sensitive to errors in the estimated statistics. Therefore, we do not use Algorithm I  for threshold adaptation. Note that the implementation complexity of the BP algorithm does not increase significantly by using Algorithm I since the channel statistics change slowly compared to the rate at which the spectrum sensing is conducted. In other words, Algorithm I is executed far less frequently than is the spectrum sensing. Note that the spectrum sensing is performed at every time slot.  

\section{Numerical Results} \label{sec:simulations}
\begin{figure}[]
%\hspace{-0.8 cm}
	  \centering 
  \includegraphics[scale=0.32]{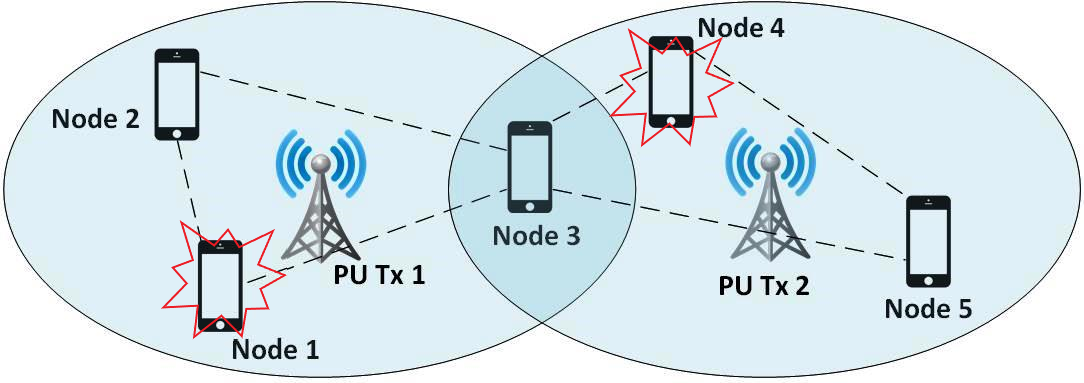} 
    \caption{We have five secondary users cooperating via BP to sense the radio spectrum allocated to a primary network with two transmitters. We use dashed lines to depict the links between the cooperating nodes, through which the BP messages are exchanged. Nodes 1 and 4 act as faulty nodes in the second experiment.}
    \label{fig:wsn} 
\end{figure}

Our simulation scenario in this section is an extension of the one considered in \cite{Abdi20}. We consider a spectrum sensing scenario typically used in cognitive radio networks \cite{Akyildiz11}. Specifically, we have five sensing nodes, as secondary users, cooperating with each other via BP to find spectral opportunities not in use by the primary users. Fig. \ref{fig:wsn} depicts our network configuration where the range of primary transmitter 1 covers nodes 1, 2, and 3 while the range of primary transmitter 2 covers nodes 3, 4, and 5. We use dashed lines to represent the links between the cooperating nodes.

%----------------Figure1---------------------------------
%---------------------------------------------------------- 

Each node generates its local sensing outcome by using energy detection while processing 100 samples of the received signals. Node 1 and node 5 receive the primary-user signal with an SNR level of -5 dB, node 2 and node 4 experience an SNR level of -8 dB in the received primary-user signal and node 3 receives signals from both of the primary transmitters at -10 dB each. In our simulations we randomly switch the primary transmitters on and off. We realize these on-off periods by generating correlated binary random variables. Hence, the primary transmitters exhibit correlated random behavior in our simulations. This is an extension to the primary network behavior assumed in \cite{Penna12}. In that work, on of the primary transmitters is on while  the other one is off and they do not change their status. As in \cite{Penna12, Abdi20}, we assume that the channel coefficients are fixed during a time slot. 

\begin{figure}[]
%\hspace{-0.8 cm}
	\centering 	
	  \includegraphics[scale=0.32]{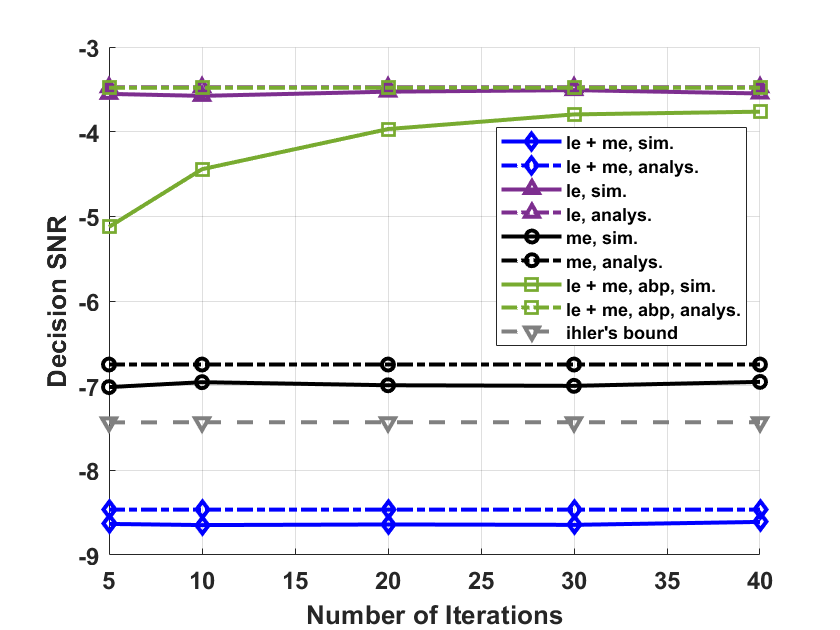} 
  \caption{Impact of errors on the decision variables built by the BP and ABP algorithms.}
  \label{fig:DSNRs} 
\end{figure}

We conduct two experiments in this section. In the first experiment, whose results are depicted in Fig.  \ref{fig:DSNRs} , we evaluate our analysis and compare its results to the one obtained by Ihler \emph{et. al.} in \cite{Ihler05}. In particular, we compare the levels of what we define as the \emph{decision SNR} (DSNR), predicted by our analysis, against the one predicted by the work in \cite{Ihler05}. We define the DSNR level at node $j$ as 
\begin{equation}
\rho_{\textup D}^{(j)} \triangleq \frac{E\left [ |\lambda_j|^2  \right ] }{E \left [|\tilde\lambda_j - \lambda_j|^2 \right]}
\end{equation}
This parameter measures the ratio of the power of the decision variable built by an error-free BP to the power of the error affecting the same decision variable in an erroneous BP. 

%----------------Figure1---------------------------------
%---------------------------------------------------------- 

%----------------Figure1---------------------------------
\begin{figure}[]
%\hspace{-0.8 cm}
	  \centering 
  \includegraphics[scale=0.32]{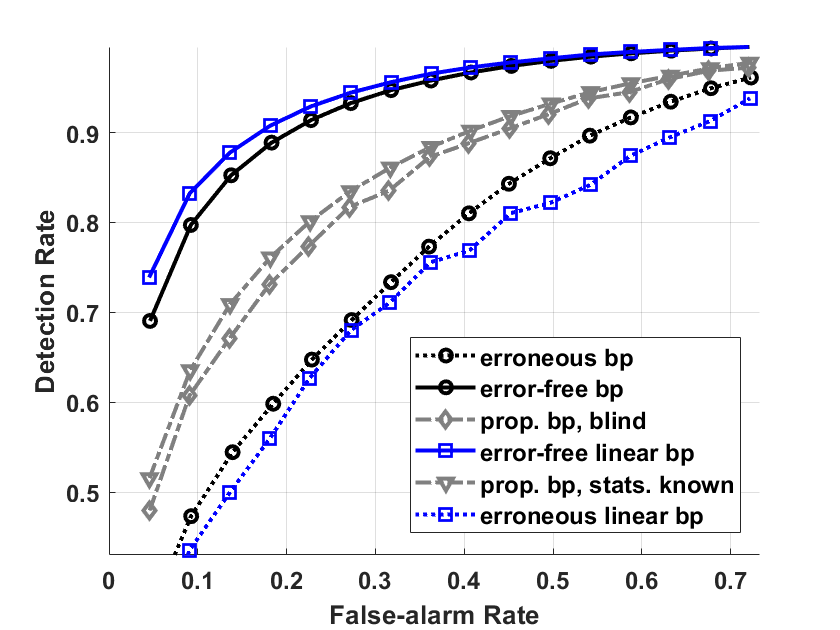} 
  \caption{Performance levels of BP, linear BP, and the proposed optimal linear BP in the presence of errors. }
  \label{fig:roc} 
   \end{figure}
%---------------------------------------------------------- 

We realize the LEs and MEs as uncorrelated zero-mean Gaussian random variables in our simulations. At each node, we measure the strength of LEs and MEs with respect to the node's local likelihood. Specifically, we define the \emph{LE SNR} level at node $j$ as 
\begin{equation}
\rho_{\textup {LE}}^{(j)} \triangleq \frac{E\left [ |\gamma_j|^2  \right ] }{E \left [|\tilde\gamma_j - \gamma_j|^2 \right]}
\end{equation}
while we define the \emph{ME SNR} level at node $j$ and for $i \in \mathcal N_j$ as 
\begin{equation}
\rho_{\textup {ME}}^{(j,i)} \triangleq \frac{E\left [ |\gamma_j|^2  \right ] }{E \left [|\tilde m_{j \to i} - m_{j \to i}|^2 \right]}
\end{equation}
Consequently, now we have a reference level at each node to measure the power of LEs and MEs injected by that node into the BP iteration. 

To see the impact of different error types separately as well as together, we run three BP algorithms in the first experiment. The first one is affected only by LEs, the second one is  affected only by MEs and the third one is affected by both of the error types concurrently. Moreover, we evaluate the behavior of ABP in the presence of both error types. In each case, the average of the DSNR level predicted by the proposed analysis and the one observed in simulations are depicted in Fig. \ref{fig:DSNRs}. The average of the DSNR is calculated over all sensing nodes in the network. The dashed curves in Fig. \ref{fig:DSNRs} represent the results of our analysis while the solid curves show the average DSNR levels observed in simulations. In this experiment, we consider $\rho_{\textup {LE}}^{(j)} = \rho_{\textup {ME}}^{(j,i)} = 10$ dB for all $i,j$ and $\zeta = 1$. In Fig. \ref{fig:DSNRs} for each data point we have averaged 20,000 realizations of the decision variables. For the adaptation process in Algorithm I we have used 2,000 data samples. That is $T = 2,000$ in this experiment. 

In Fig. \ref{fig:DSNRs} we see a close match between the results predicted by our analysis and the ones obtained via simulations. We also see that our analysis provides a better estimate of the DSNR levels than Ihler's bound in \cite{Ihler05}. \emph{Our analysis provides a better estimation even when we only consider MEs}. 

Moreover, we see a gap in Fig. \ref{fig:DSNRs} between the DSNR levels of the LE-only and ME-only cases. This gap indicates that the impact of MEs is more deteriorating than the impact of LEs when we have the same levels of $\rho_{\textup{LE}}$ and $\rho_{\textup{ME}}$. This appears to be a reasonable observation in our network since the set of error items injected to the BP iteration by each node, say node $j$, comprises of only one LE and $|\mathcal N_j|$ MEs and $|\mathcal N_j| \ge 1$. This observation can be justified based on the linearity of the proposed detection scheme. Specifically, the average number of neighbors in our network is $(4\times2 + 4)/5 = 2.4$ and $10\log_{10}(2.4) \approx 3.8$ dB and we see almost 3.5 dB gap between the two curves. Note that Ihler's bound  \eqref{eq:E_ln_d} cannot distinguish between the LEs and MEs.

Fig. \ref{fig:DSNRs} confirms our observation in Sec. \ref{subsec:Ave} where we showed that ABP is quite resilient to the impact of MEs. We now see that, by increasing the number of iterations the average DSNR level of ABP approaches that of a BP that is affected by LEs only. Note that the ABP is affected by both types of errors and even when the number of iterations is rather low, the DSNR level of ABP is quite high compared to that of a regular BP affected by both error types. This observation justifies our choice of ABP for the offline learning and optimization cycle in Algorithm \ref{tab:offlineBP}. 

In this experiment, for a given number of iterations $N_{\textup{iterate}}$, we set $L$ large enough, i.e., $L \ge N_{\textup{iterate}}$, to use all the messages generated by BP when realizing the averaging process in the ABP decision variable \eqref{eq:m_kj_bar}. Hence, by increasing the number of iterations, $L$ is increased and this increase leads to a heavier suppression of MEs. We can clearly see in Fig. \ref{fig:DSNRs} the increase in the DSNR level of ABP caused by increasing $N_{\textup{iterate}}$. Moreover, the DSNR level of the ABP approaches that of LE-only case, which, as we predicted in Sec. \ref{subsec:Ave}, indicates that the LEs are not affected by the averaging process in the ABP. 

In addition, our analysis, which is based on the von Neumann model, predicts that the DSNR levels of an erroneous BP algorithm do not change by the number of iterations significantly. This is also confirmed by the simulation results in Fig. \ref{fig:DSNRs}. 

In the second experiment, we study the impact of errors on the detection performance of the sensor network depicted in Fig. \ref{fig:wsn}. We assume that nodes 1 and 4 are faulty and inject errors into the BP algorithm. All other nodes operate in a reliable manner, meaning that the errors in their local likelihoods and messages are negligible. The results are depicted in Fig. \ref{fig:roc} where for each data point we have averaged 100,000 detection results. For the adaptation of the linear BP we have used a window of 2500 detection outcomes. That is $T = 2500$ in Algorithm I. We have set $L = 10$ in the offline ABP of Algorithm \ref{tab:offlineBP}. In the faulty nodes we have $\rho_{\textup {LE}}^{(1)} = \rho_{\textup {LE}}^{(4)} = 10$ dB while $\rho_{\textup {ME}}^{(1)} = \rho_{\textup {ME}}^{(4)} = 20$ dB. The aim of this experiment is to see whether the proposed method is able to alleviate the impact of those faulty nodes on the overall detection performance. 

As for performance metrics, we use the average of the detection and false-alarm rates observed in all of the sensing nodes. We consider both the BP and linear BP algorithms with error-free and erroneous iterations. Consequently, we see how each detection method is affected by errors. Moreover, we consider the proposed linear BP optimized with and without having the required statistics. Fig. \ref{fig:roc} shows that, in the presence of LEs and MEs, the detection performance of both message-passing algorithms are significantly degraded. This observation clarifies the need for a better BP algorithm which resists against the impact of errors. In Fig. \ref{fig:roc} we also see that the proposed method significantly improves the detection rate of the system in the presence of errors. Moreover, we can see that the proposed blind adaptation scheme closely achieves the optimal performance level when the required statistics are not available a priori.

\section{Conclusion} \label{sec:conclusions}
We studied the impact of computation and communication errors on the behavior of the BP algorithm. We showed that when evaluating the impact of errors on a distributed detection conducted by BP, the detection can effectively be modeled as a distributed linear data-fusion scheme. Consequently, we can analyze its statistical behavior in the presence of errors and obtain closed-form relations for its performance metrics. Moreover, by optimizing the resulting linear data-fusion we can effectively suppress the impact of errors and obtain a better detection performance. 

\section*{Appendix I: Proof of \eqref{AandC}} \label{appendix}
We focus on non-diagonal elements of $\boldsymbol{A}$ here since it is clear that the diagonal ones are all close to one. It is straightforward to see that \eqref{AandC} holds for $l = 1,2,3$. That is, we have $\boldsymbol{\lambda}^{(l)} = \boldsymbol{A}^{(l)}\boldsymbol{\gamma}$ where $\boldsymbol{A}^{(l)} \approx  \sum_{n = 1}^{l}\boldsymbol{C}^n$ for $l \le 3$. Based on this observation, we prove \eqref{AandC} by induction. Specifically, we show that if $\boldsymbol{A}^{(k)} \approx \sum_{n = 1}^{k}\boldsymbol{C}^n$ for $k \le l$ then $\boldsymbol{A}^{(l +1)} \approx  \sum_{n = 1}^{l + 1}\boldsymbol{C}^n$ or, equivalently, $\boldsymbol{A}^{(l+1)} \approx \boldsymbol{A}^{(l)} + \boldsymbol{C}^{l + 1}$. From \eqref{lambda_j_2} and \eqref{m_kj_linear} and by setting $c_{jk} = 0$ for $k \notin \mathcal{N}_j, \forall j$ we have
\begin{align}\label{A(l+1)}
&\left [\boldsymbol{A}^{(l+1)}  \right ]_{ji} = \left [\boldsymbol{A}^{(l)}  \right ]_{ji} \nonumber \\
&+ \sum_{k_1 \neq j} \sum_{k_2 \neq j} ... \sum_{k_{l-1} \neq k_{l-3}}\sum_{k_l\neq k_{l-2}} c_{jk_1}c_{k_1k_2} ... c_{k_{l-1} k_l}c_{k_l i}
\end{align}
Note that $ \left [\boldsymbol{A}^{(l)}  \right ]_{ji} = {\partial \lambda_j/\partial \gamma_i}$. Hence, to prove \eqref{AandC} we need to show that 
\begin{equation} \label{C(l+1)}
\left [\boldsymbol{C}^{l+1} \right ]_{ji} \approx \sum_{k_1 \neq j} \sum_{k_2 \neq j} ... \sum_{k_l\neq k_{l-2}}  c_{jk_1}c_{k_1k_2} ... c_{k_l i}
\end{equation}
According to \eqref{A(l+1)}, our induction hypothesis indicates for $n \le l$ that 
\begin{equation} \label{eq:C(l)}
\left [\boldsymbol{C}^{n}   \right ]_{ji} \approx \sum_{k_1 \neq j} \sum_{k_2 \neq j} ... \sum_{k_{n-1} \neq k_{n-3}} c_{jk_1}c_{k_1k_2} ... c_{k_{n-1} i}
\end{equation}
To show that \eqref{C(l+1)} is true, we rewrite its right-hand side (RHS) as 
\begin{align}\label{eq:rhs}
&\textup{RHS} = \sum_{k_l = 1}^N c_{k_l i}  \sum_{k_1 \neq j} \sum_{k_2 \neq j} ... \sum_{k_{l-1} \neq k_{l-3}} c_{jk_1}c_{k_1k_2} ... c_{k_{l-1} k_l}  \nonumber \\
	&- \sum_{k_1 \neq j} \sum_{k_2 \neq j} ... \sum_{k_{l-1} \neq k_{l-3}} c_{jk_1}c_{k_1k_2} ...  c_{k_{l-2} k_{l-1}}c_{k_{l-1} k_{l-2}}c_{k_{l-2} i} 
\end{align}
which, based on \eqref{eq:C(l)}, means that 
\begin{equation}\label{C(l+1)_short}
\textup{RHS} \approx \sum_{k_l = 1}^N c_{k_l i}  \left [\boldsymbol{C}^{l}   \right ]_{jk_l} - O_j^{(l+1)}
\end{equation}
where $O_j^{(l+1)}$ denotes the right-most term in \eqref{eq:rhs}. 

Now, we recognize the outcome of the first sum in \eqref{C(l+1)_short} as $\left [\boldsymbol{C}^{l+1} \right ]_{ji}$ which is what we were looking for. Since the induction hypothesis states that $\boldsymbol{A}^{(l)}$ contains $\boldsymbol{C}^{l-1}$ as one of its summands, the offset term $O_j^{(l+1)}$ can be neglected. The reason is that 
\begin{equation} \label{eq:upBound}
\left |{O_j^{(l+1)} \over \left [\boldsymbol{C}^{l-1}   \right ]_{ji} } \right| \le \left |\left (|\mathcal{N}_{k_{l-1}}|-1  \right )\tilde{c}^2 \right |
\end{equation}
where $\tilde{c} \triangleq {1 \over \max_n|\mathcal{N}_n|-1} <1$ and, therefore, $|\tilde{c}|^2 \ll 1$. The bound in \eqref{eq:upBound} is derived by replacing the term $c_{k_{l-2} k_{l-1}}c_{k_{l-1} k_{l-2}}$ within $O_j^{(l+1)}$ by $\tilde{c}^2$. Recall that, to ensure the system convergence we have $|c_{j,k}| < \tilde{c}, \forall (j,k) \in \mathcal{E}$. Hence, the proof is complete.

\bibliographystyle{IEEEtran}
\bibliography{IEEEabrv,Bibliogeraphy_ICLAS}

\end{document}